\documentclass[review]{elsarticle}
\usepackage{amsmath}
\usepackage{amssymb}
\usepackage{lineno,hyperref}
\usepackage{color}

\usepackage[percent]{overpic}
\usepackage{cleveref}
\usepackage{multirow}
\usepackage{booktabs}
\crefname{equation}{Eq.~}{Eqs.~}
\crefname{figure}{Fig.~}{Figs.~}
\crefname{section}{section}{sections}
\journal{ Results in physics }
\begin{document}

\begin{frontmatter}

\title{Parametrized protocol achieving the Heisenberg limit in the optical domain via dispersive atom-light interactions}

\author[mymainaddress]{{Yuguo Su}}
%

\author[mymainaddress]{Xiaoguang Wang\corref{mycorrespondingauthor}}
\cortext[mycorrespondingauthor]{Corresponding author}
\ead{xgwang1208@zju.edu.cn}

\address[mymainaddress]{Zhejiang Province Key Laboratory of Quantum Technology and Device, Department of Physics, Zhejiang University, Hangzhou, Zhejiang 310027, China}

\begin{abstract}
The strong and collective atom-light interactions in cavity-QED systems perform manifold benefits in quantum-enhanced measurements.
Here, we study the time-reversal protocol that has been proposed to sense small displacements of the light field, and report the sensitivity of the scheme that could be speeded up to attain the Heisenberg limit (HL).
We show the holonomic unitary parametrization process of the scheme and one only need to choose appropriate initial states to pursue the ultimate sensitivity.
The scheme may pave an experimentally feasible way to achieve Heisenberg-limited metrology with nonclassical states.
\end{abstract}

\begin{keyword}
Quantum metrology, Cavity quantum electrodynamics, Unitary parametrization process, Quantum sensing
\end{keyword}

\end{frontmatter}


\section{Introduction}\label{Sec. I}
Quantum metrology, which encompasses the estimation of an unknown parameter of a quantum system, is of  importance in fundamental physics and technology.
With different models of the probe systems and different parameters to be estimated, many applications of quantum metrology have been done
and emerged manifold potential in quantum technology such as optimal quantum clock \cite{Bufmmodeheckzlseziek1999}, clock synchronization \cite{Jozsa2000}, measurement of gravity accelerations \cite{Peters1999} and quantum frequency standards \cite{Bollinger1996}.
It is well known that using quasiclassical states the sensitivity is bounded by the standard quantum limit (SQL) \cite{Caves1981,Wineland1992}.
However, the measurement precision of the parameter is constrained by a fundamental limit, the quantum Cram{\'e}r-Rao bound (QCRB) \cite{Braunstein1994} $\Delta\phi\geq1/\sqrt{\mathcal{F}}$, where $\mathcal{F}$ is the quantum Fisher information (QFI).
For $N$ particles in each shot, the SQL, also known as the shot-noise limit, can be surpassed by using quantum effects such as entanglement \cite{Bollinger1996,Monz2011} and squeezing \cite{Muessel2015}, reaching the so-called Heisenberg limit (HL), in which the sensitivity exceeds the SQL by $1/\sqrt{N}$ \cite{Bollinger1996,Holland1993,Munro2002}.
Many schemes have been proposed to achieve the SQL, such as quantum state transfer from light to atoms \cite{Agarwal1990,Kuzmich1997,Moore1999}, quantum nondemolition measurement \cite{Appel2009,Kuzmich1998,Louchet-Chauvet2010,Hammerer2010}, one-axis twisting \cite{Schleier-Smith2010,Kitagawa1993,Sorensen2001,Haine2014}, two-axis countertwisting \cite{Kitagawa1993,Ma2009}, twist-and-turn squeezing \cite{Muessel2015,Law2001}, spin changing collisions \cite{Liicke2011,Duan2000,Pu2000,Nolan2016}, and adiabatically scanning through a quantum phase transition \cite{Lee2006,Huang2018}.
Furthermore, the concept of interaction-based readout resolves the dilemma that the states prepared via these schemes require the low-noise detection in order to see significant quantum enhancement \cite{Haine2018,Demkowicz-Dobrzanski2012}.

The field of semiconductor cavity-QED has become the rapidly growing subject of intense theoretical and experimental studies due to the potential applications in quantum information science \cite{Reithmaier2004,Yoshie2004}.
The field of cavity-QED exhibits the interactions between electromagnetic fields in cavities and electronic states of matter \cite{wang_turning_2019,Pino2015,zhang_photon-assisted_2020}.
It offers unprecedented scientific possibilities for exploring various applications in quantum computing \cite{Bennett2000,Knill2001,Buluta2010}, quantum key distribution \cite{Scarani2009,OBrien2009}, quantum simulation \cite{Buluta2009,Georgescu2014}, and quantum metrology \cite{Giovannetti2011,zhang_quantum_2018}.
As a precedent, one possible application associated with the vibrational displacement interactions in cavity-QED systems is to generalize entangled coherent states \cite{Sanders2012,Sanders1992} and superposed coherent states \cite{Milburn1986,Milburn1986a,Yurke1986,Armour2002,Marshall2003,Liao2016,Liao2016a}.
Nowadays, the development of quantum-enhanced sensors in the cavity-QED attracts attention \cite{Lewis-Swan2020,barberena2020,Norcia2018,Safavi-Naini2018,Cox2016,Norcia2016,Bohnet2015}, and many proposals for the generation of entangled coherent states have been investigated in many physical systems, including cavity-QED systems \cite{Kim1999,Solano2003,Akram2013}, atomic-BEC systems \cite{Kuang2003} and ion traps \cite{Gerry1997,Munro2000}.

The spin-coherent state (SCS), as known as the atomic coherent state, is employed to denote the states where the collective spin has the minimum uncertainty in an ensemble of $N$ spin-$J$ particles \cite{Arecchi1972}.
The SCSs, which are product states and no quantum entanglement is presented between the qubits, have been applied in many branches of physics.
The SCSs can be widely harnessed in modern quantum technologies such as quantum metrology \cite{Solano2003,Joo2012,Su2020PRA} and quantum teleportation based on coherent-state bases \cite{Wang2001,Enk2001,Johnson2002}.
The ability to access an SCS of entangled spins also exploits up a new domain for the study of the interrelation between collective spin behaviors and quantum correlations in a spin ensemble \cite{Sorensen2001,Guehne2009}.

In quantum metrology, the strong and collective atom-light interactions in cavity-QED systems exhibit prominent advantage in quantum-enhanced measurements.
In this work, we study the time-reversal protocol to sense small displacements of the light field, and show the sensitivity of the scheme which could be speeded up to attain the HL.
We elucidate the holonomic unitary parametrization process of the scheme and one only need to choose appropriate initial states to seek the ultimate sensitivity.
Our scheme may provide new experimental possibilities to achieve Heisenberg-limited metrology with nonclassical states.

This paper is organized as follows. In ~\cref{Sec. II}, we concretely introduce the steps of the time-reversal protocol.
We show its unitary parametrization process and obtain the expression of the QFI.
Furthermore, we gain the sensitivities of the small displacements of the light field by choosing the optical part state as superposition of even and odd coherent state, and changing the atomic part state from the collective ground state to the superposed spin-coherent state in~\cref{Sec. III} and~\cref{Sec. IV}.
A summary is given in ~\cref{Sec. V}.

\section{Time-Reversal Protocol}\label{Sec. II}
Nowadays, a time-reversal protocol has been demonstrated as a feasible solution to sense a small perturbation of a phase-space structure state, for example, the cat states.
The time-reversal protocol, wherein the initial entangling dynamics are reversed after the perturbation, reconciles the problem we meet: the measurements of phase-space structure states entail the feasibility of state tomography \cite{Toscano2006,Zurec2001} or single-particle resolution \cite{Bollinger1996}.
Furthermore, reversal protocols prossess other striking merits encompassing high-feasibility in experiments and robustness to experimental detection
noise \cite{Davis2016,Nolan2017,Haine2018,Mirkhalaf2018}.

\begin{figure}
	\centering
	\begin{minipage}{0.75\linewidth}
		\begin{overpic}[width=\linewidth]{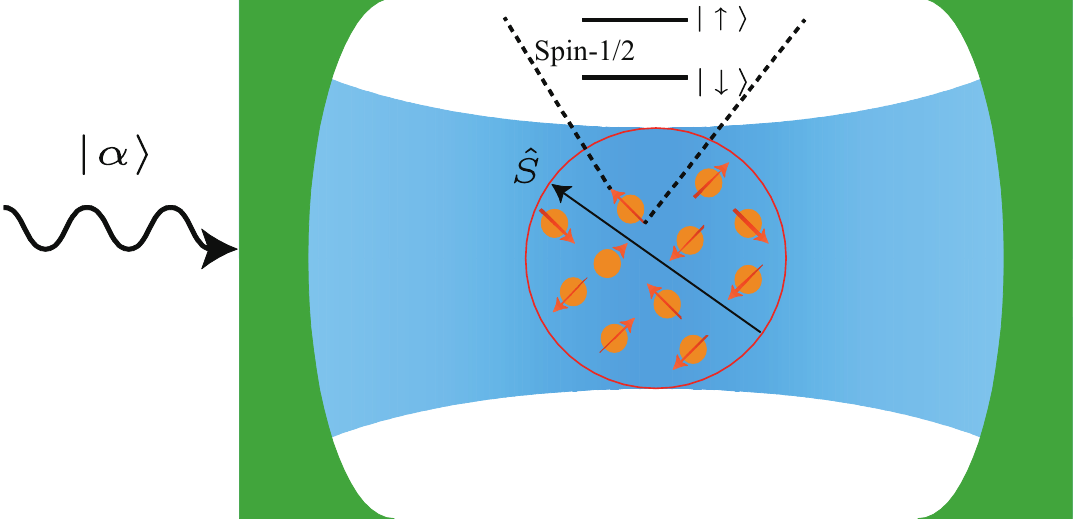}
						\put(0,50){$(a)$}
		\end{overpic}
	\end{minipage}
	\begin{minipage}{1\linewidth}
		\begin{overpic}[width=\linewidth]{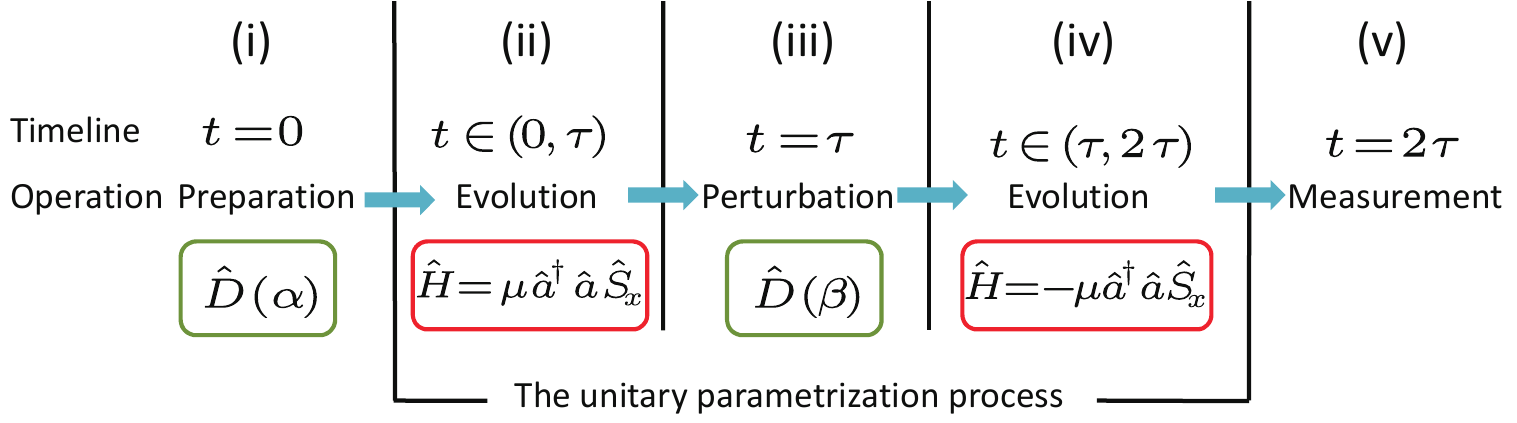}
					\put(0,25){$(b)$}
		\end{overpic}
	\end{minipage}
	\protect\protect\caption{
		(a) Cavity-QED setup. A coherent field $\alpha$ is injected into the cavity to which a collective spin $\hat{S}$ comprised of $N$ spin-1/2 particles is coupled.
		(b) Schematic depiction of the protocol.
		More details could be found in~\cite{Lewis-Swan2020,barberena2020}. 
	}
	\label{fig1}
\end{figure}

To parametrize the interferometric protocol proposed in~\cite{Lewis-Swan2020,barberena2020}, we introduce the interferometric protocol that consists of the following steps (\cref{fig1}(a) and~(b), more details seen in~\cite{Lewis-Swan2020,barberena2020}): (i) prepare the cavity in a coherent state of the real amplitude $\alpha$ and with an $N$ spin ensemble (pseudospin) pointing along $-\hat{S}_z$, (ii) evolve with $\hat{H}$ during time $t\in\left(0,\tau\right)$, (iii) coherently impose a perturbation $\beta$ of the displacement on the cavity at time $t=\tau$, (iv) evolve with $-\hat{H}$ during time $t\in\left(\tau,2\tau\right)$, and (v) implement a measurement at final time $t=2\tau$.

From~\cite{Blais2004,2007Resolving}, the dispersive atom-light coupling Hamiltonian is given by
\begin{equation}
\hat{H}=\mu\hat{a}^\dagger\hat{a}\hat{S}_x,
\end{equation}
where $\mu$ is the atom-light coupling constant, $\hat{a}$ ($\hat{a}^\dagger$) are annihilation (creation) operators for the cavity mode, $\hat{S}_{x,y,z}=\sum^{N}_{j=1}{\hat{\sigma}^{j}_{x,y,z}/2}$ are the collective spin operators with Pauli operators $\hat{\sigma}^{j}_{x,y,z}$ acting on atom $j$, $N$ is the number of spins, and we set $\hbar=1$ throughout this paper.
To parametrize the protocol during the whole evolution $t\in\left(0,2\tau\right)$, we gain the unitary operator of this system $\hat{U}=e^{i\theta \hat{a}^{\dagger}\hat{a}\hat{S}_{x}}\hat{D}\left(\beta\right)e^{-i\theta \hat{a}^{\dagger}\hat{a}\hat{S}_{x}}$, where $\theta=\mu t$ is a time-dependent phase factor and $\hat{D}=e^{\beta \hat{a}^{\dagger}-\beta^{*}\hat{a}}$ is the displacement operator.
We can employ the QFI to quantify the sensitivity and it can be expressed as \cite{Boixo2007,Taddei2013,Liu2015}
$\left.\mathcal{F}=4\left\langle\Delta^{2}\hat{\mathcal{H}}\right\rangle=4\left(\left\langle \hat{\mathcal{H}}^{2}\right\rangle -\left\langle \hat{\mathcal{H}}\right\rangle ^{2}\right)\right.$, where $\hat{\mathcal{H}}=i\left(\partial_{\beta}\hat{U}^{\dagger}\right)\hat{U}$ is a Hermitian generator with respect to the estimated real parameter $\beta$ and is independent of the initial state.
Due to the intrinsic property of the QFI, the optimal QFI is independent of the specific measurements.
Moreover, since the Hermitian generator $\hat{\mathcal{H}}$ absorbs all the information of the unitary parametrization process, we can focus on choosing the initial state of the time-reversal protocol.

Then after some algebraic calculation, the QFI of the time-reversal protocol could be gained as (see Appendix~A for detailed derivations)
\begin{equation}
\mathcal{F}  = \!4\!\left[\!2\left\langle \! \hat{a}^{\dagger}\!\hat{a}\!\right\rangle \! +\!1\!-\!\left(\!\left\langle \!\hat{a}^{\dagger2}\right\rangle \!\left\langle \!e^{2i\theta \hat{S}_{x}}\!\right\rangle \!+\!\left\langle \!\hat{a}^{2}\right\rangle \!\left\langle \!e^{\!-\!2i\theta \hat{S}_{x}}\!\right\rangle \!\right)
\!+\!\left(\!\left\langle \!\hat{a}^{\dagger}\right\rangle \!\left\langle \!e^{i\theta \hat{S}_{x}}\!\right\rangle \!-\!\left\langle \hat{a}\right\rangle \!\left\langle \!e^{\!-i\theta \hat{S}_{x}}\!\right\rangle \!\right)^{2}\right].\label{F_1}
\end{equation}
It is clear that the QFI consists of the optical contribution and the atomic contribution.


\section{Superposition of even and odd coherent state, and collective ground state}\label{Sec. III}
In this section, we consider that the initial state is the product state of the even and odd coherent superposition state, and the collective ground state (the case-a).
The superposition of even and odd coherent state is defined as \cite{Ansari1994,Dodonov1995}
\begin{eqnarray}
\left|\psi\right\rangle _{opt}=\frac{1}{\sqrt{2\left(1+e^{-2\alpha^{2}}\cos\phi\right)}}\left(\left|\alpha\right\rangle +e^{i\phi}\left|-\alpha\right\rangle \right),
\end{eqnarray}
and one can obtain the relevant average values
\begin{eqnarray}
\left\langle \hat{a}\right\rangle & =&\frac{-i\,\alpha \,e^{-2\left|\alpha\right|^{2}}\sin\phi}{1+e^{-2\left|\alpha\right|^{2}}\cos\phi},\\
\left\langle \hat{a}^{2}\right\rangle  &
=&\alpha^2,\\
\left\langle \hat{a}^{\dagger}\hat{a}\right\rangle  & =&\frac{\left|\alpha\right|^{2}\left(1-e^{-2\left|\alpha\right|^{2}}\cos\phi\right)}{1+e^{-2\left|\alpha\right|^{2}}\cos\phi},
\end{eqnarray}
where $\phi$ is a relative phase factor.
If we take $\alpha=\alpha^{*}$, then we have the QFI
\begin{eqnarray}
\mathcal{F}  &=&  4+4\alpha^{2}\left[\frac{2\left(1-e^{-2\alpha^{2}}\cos\phi\right)}{1+e^{-2\alpha^{2}}\cos\phi}-\left(\left\langle e^{2i\theta \hat{S}_{x}}\right\rangle +\left\langle  e^{-2i\theta \hat{S}_{x}}\right\rangle  \right)\right.-\nonumber\\
&&\left.\left(\frac{e^{-2\alpha^{2}}\sin\phi}{1+e^{-2\alpha^{2}}\cos\phi}\right)^{2}\left(\left\langle e^{i\theta \hat{S}_{x}}\right\rangle +\left\langle e^{-i\theta \hat{S}_{x}}\right\rangle \right)^{2}\right].
\end{eqnarray}
In contrast, if we choose the Fock state as the optical part state, we find that only the mean photon number $\left\langle \hat{a}^{\dagger}\hat{a}\right\rangle $
survives and the QFI is $\mathcal{F}=4+8\left\langle \hat{a}^{\dagger}\hat{a}\right\rangle $.

For the atomic part state, we employ the collective ground state of $\hat{S}_z$
\begin{eqnarray}
\left|\psi\right\rangle _{atm} & =\left|\downarrow\right\rangle ^{\otimes N}=\left|0\right\rangle,
\end{eqnarray}
where $\left|n\right\rangle\equiv \left|N/2,n-N/2\right\rangle$ is the Dicke state of the collective spin and $n=0,1,2,\dots$.
To calculate the atomic contributions, we introduce the rotation operator, defined formally as \cite{Arecchi1972}
\begin{eqnarray}
\hat{R}\left(\vartheta,\varphi\right) & =&  e^{i\vartheta\left(\hat{S}_{x}\sin\varphi-\hat{S}_{y}\cos\varphi\right)}\label{R1}\\
& = & e^{-\eta^{*}\hat{S}_{-}}e^{-\ln\left(1+\left|\eta\right|^{2}\right)\hat{S}_{z}}e^{\eta \hat{S}_{+}}\label{R2}\\
& =&  e^{\eta \hat{S}_{+}}e^{\ln\left(1+\left|\eta\right|^{2}\right)\hat{S}_{z}}e^{-\eta^{*}\hat{S}_{-}},\label{R3}
\end{eqnarray}
which produces a rotation through an angle $\vartheta$ about an axis $\hat{n}=\left(-\sin\varphi,\cos\varphi,0\right)$ in the Bloch sphere,
where $\vartheta$ is the zenith angle and $\varphi$ is the azimuthal angle and $\eta=\frac{\varsigma}{\left|\varsigma\right|}\tan\left|\varsigma\right|=-e^{-i\varphi}\tan\frac{\vartheta}{2}$, $\varsigma=-\frac{\vartheta}{2}e^{-i\varphi}$.
Utilizing the rotation operator to the collective ground state, one can gain
\begin{eqnarray}
e^{-i\theta \hat{S}_{x}}\left|0\right\rangle  =  \hat{R}\left(\theta,\frac{\pi}{2}\right)\left|0\right\rangle  =  \cos^{N}\frac{\theta}{2}\sum_{k=0}^{\infty}\frac{\left(-i\tan\frac{\theta}{2}\right)^{k}}{k!}\frac{N!}{\left(N-k\right)!}\left|k\right\rangle .
\end{eqnarray}
Consequently, we have the atomic contributions $\left\langle e^{\pm i\zeta\theta \hat{S}_{x}}\right\rangle =\cos^{N}\left(\zeta\theta/2\right)$
and the QFI of the time-reversal protocol
\begin{equation}
\mathcal{F}  \!=\!4\!+\!8\alpha^{2}\!\left[\frac{e^{4\alpha^{2}\!}\!-\!2\!\cos^{2N}\!\frac{\theta}{2}\sin^{2}\!\phi\!-\!\cos^{2}\!\phi\!}{\left(e^{2\alpha^{2}}+\cos\phi\right)^{2}}\!-\!\left(\!2\cos^{2}\!\frac{\theta}{2}\!-\!1\!\right){}^{\!N\!}\right],\label{F_2}
\end{equation}
where $\zeta=1,\,2$ is a constant.

Through some calculations (see Appendix~B for detailed analyses), we find that if $y\geq0$ or $\alpha\in\left[\alpha_{1},\infty\right)$, the QFI $\mathcal{F}$ is monotonically increasing (decreasing) when $\phi\in\left[2k\pi,\pi+2k\pi\right)$ ($\phi\in\left[\pi+2k\pi,2\pi+2k\pi\right)$); if $\left.y<0\right.$ and  $\left.\alpha\in\left[0,\alpha_{1}\right)\right.$, the QFI $\mathcal{F}$ is monotonically increasing (decreasing) when $\phi\in\left[\pi+2k\pi,2\pi+2k\pi\right)$ ($\phi\in\left[2k\pi,\pi+2k\pi\right)$), where $\left.\!y=\!1-2\cos^{2N}\left(\theta/2\right)\right.$, $k\in\mathbb{Z}$ and $\alpha_1= \! \sqrt{\ln\left\{ \left[\!-y\cos\!\phi\!+\!\sqrt{4\left(1-y\right)+y^2 \cos^{2}\!\phi}\right]/2\right\}/2 }$.

Therefore, we obtain the general expression of  the maximal QFI 
\begin{equation}
\mathcal{F}_{max}=
\begin{cases}
\left.\mathcal{F}\right|_{\phi=\pi+2k\pi}, &  y \geq0\ or\ \alpha>\alpha_{1},\\
	\left.\mathcal{F}\right|_{\phi=2k\pi},& y<0\ and\ 0<\alpha<\alpha_{1},
\end{cases}
\end{equation}
and the maximal one with respect to the relative phase factor $\phi$ 
\begin{eqnarray}
\mathcal{F}_{max} & =  4+8\alpha^{2}\!\left[\frac{e^{4\alpha^{2}}\!-\!1}{\left(e^{2\alpha^{2}}\!-\!1\right)^{2}}\!-\!\left(2\cos^{2}\frac{\theta}{2}-1\right){}^{N}\right],
\end{eqnarray}
when the optical part of the initial state is an odd state ($\phi=\pi+2k\pi$).

\begin{figure}[th]
	\centering
	\includegraphics[scale=1]{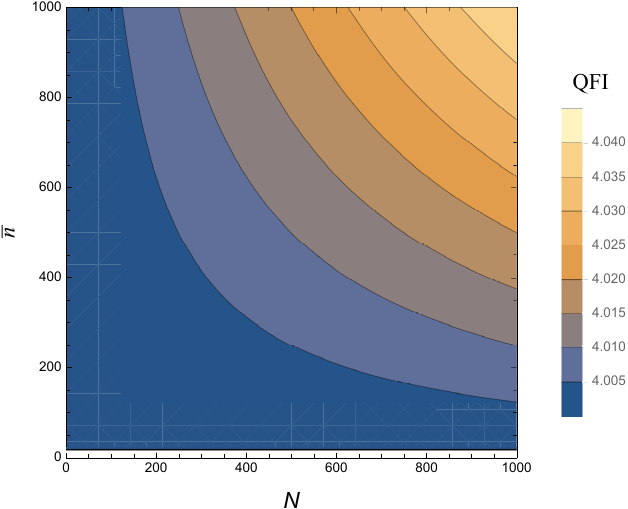}
	\protect\caption{
		Maximal QFI of the product state of the even and odd coherent superposition state and the collective ground state.
		The contour map of the QFI versus the spin number $N$ and the mean photon number $\bar{n}$ at short time $\tau=10^{-4}/\mu$.
	}\label{fig2}
\end{figure}

For short times $\tau\ll1/\left(\mu\sqrt{N}\right)$, which means 
\begin{eqnarray}
\lim_{\theta\rightarrow0}\left\langle e^{\pm i\zeta\theta \hat{S}_{x}}\right\rangle  & =  \lim_{\theta\rightarrow0}\cos^{N}\left(\frac{\zeta}{2}\theta\right)\approx1-\frac{1}{8}N\zeta^{2}\theta^{2},
\end{eqnarray}
one can get the maximal QFI as
\begin{eqnarray}
\mathcal{F}_{max}\! & =& \! 4\!+\!4\alpha^{2}\!\left(\!N\theta^{2\!}\!-\!2+\!2\!\coth\!\alpha\!^{2}\right)\!\approx\!4\!+\!4N\alpha^{2}\mu^{2}\tau^{2}\!,
\end{eqnarray}
when the mean photon number is large enough [\cref{fig2}]. 
Here, the SQL is a constant, $\Delta^2\beta_{SQL}=1/4$, which is defined as the sensitivity with the original coherent state $\left|\alpha\right\rangle$.

At longer times $\tau\gtrsim1/\left(\mu\sqrt{N}\right)$, from the derivative of the QFI versus the time-dependent phase factor $\theta$ $\left.\partial_{\theta}\mathcal{F}=4\alpha^{2} N \sin\left(2\theta\right)\cos^{N-2}\theta\right.$, one can gain the maximal QFI as
\begin{equation}
\mathcal{F}_{max}=\!4+8\alpha^{2}\!\left[\!\frac{\!1\!-\!\left(\!-\!1\!\right)^{N}}{2}\!+\!\coth \!\left(\!\alpha^{2}\!\right)\!\right]\!
\approx\!\begin{cases}
	\!4\!+\!16\alpha^{2}, &\!\theta\!=\!\pi\!+\!2k\pi\ and\ odd\!-\!N,\\
	\!4\!+\!8\alpha^{2}, &\!\theta\!=\!\frac{\pi}{2}\!+k\pi\ and\ even\!-\!N,
	\end{cases}
\end{equation}
when the mean photon number is large enough. 
Hence, the odd spin number is optimal at longer times. 
The explanation of the non-spin relevance is that the atomic fluctuations will superpose the bosonic coherent state completely about a circle of radius $\left|\alpha\right|$ in phase space, and the state is equivalently sensitive to perturbations along any direction.


\section{Superposition of even and odd coherent state, and superposed spin-coherent state}\label{Sec. IV}
In this section, we consider that the initial state is the product state of the even and odd coherent superposition state and the superposed spin-coherent state (the case-b).
The optical contributions have been discussed in the previous section.
Now, we focus on the atomic contributions and introduce the spin-coherent superposition state, defined as \cite{Huang2018a,Pawlowski2013}
\begin{eqnarray}
\left|\psi\right\rangle _{atm}=\frac{1}{\sqrt{2}}\left(\left|\eta\right\rangle +e^{i \nu}\left|-\eta\right\rangle \right),
\end{eqnarray}
where the spin-coherent state is given by 
\begin{eqnarray}
\left|\eta\right\rangle \!  & = \! \left|\vartheta,\varphi\right\rangle
\!=\! \hat{R}\left(\vartheta,\varphi\right)\left|0\right\rangle 
\!=\!  \left(\!\frac{1}{1+\left|\eta\right|^{2}}\!\right)^{\!\frac{N}{2}}\!e^{\eta \hat{S}_{+}}\!\left|0\right\rangle \label{SCS},
\end{eqnarray}
$\nu$ is a relative phase factor and the rotation operator $\hat{R}\left(\vartheta,\varphi\right)$ has been
introduced in the previous section.

For short times $\tau\ll1/\left(\mu\sqrt{N}\right)$, analyzing the
expression of the atomic contributions (see Appendix~C for detailed
derivations), we have
\begin{eqnarray}
-\left\langle \! e^{-2i\theta \hat{S}_{x}}\! +\!e^{2i\theta \hat{S}_{x}}\! \right\rangle \! &=&
\!-2\!+\!N\theta^{2}\!+\!\left(N-1\right)N\theta^2 \sin^{2}\!\vartheta \cos^2 \! \varphi+\cos\nu  \cos^{N-2} \vartheta\nonumber\\
&& \times\left[\left(N\theta^2-2\right)\cos^{2} \vartheta  \!+\! \left(1-N\right)N\theta^2 \sin^2 \vartheta \cos^2 \varphi  \right],\label{at.1}
\end{eqnarray}
where $N\theta ^{2}\ll 1$, $\cos\theta\approx1-\theta^2/2$, $\sin\theta\approx\theta$ and the spin number $N$ is large enough.
Analyzing the sign of each terms of Eq.~\eqref{at.1}, one can gain the maximum with respect to the relative phase factor $\nu$
\begin{eqnarray}
-\left\langle \! e^{\!-2i\theta \hat{S}_{x}}\! +\!e^{2i\theta \hat{S}_{x}}\! \right\rangle \! & =&
2\left(\cos^N \vartheta\!-\!1\right)\!+\!N\theta^2 \left(1\!-\!\sin^2 \vartheta \cos^2 \varphi\right)\nonumber\\
&&\!-\! N\theta^2 \cos^{N-2}\vartheta \left(\cos^2 \vartheta+\sin^2 \vartheta \sin^2 \varphi\right)\nonumber\\
&& \!+\!N^2 \theta^2 \sin^2 \vartheta \!\left(\cos^2 \varphi \!+\!\sin^2 \varphi \cos^{N-2}\vartheta\right), \label{at.1.nu}
\end{eqnarray}
where $\nu=\pi+2k\pi$.
Since $N$ is large enough ($N^2\theta^2\gg1$), one know that the last term of Eq.~\eqref{at.1.nu} is dominant and obtain the maximum 
\begin{equation}
-\left\langle  e^{-2i\theta \hat{S}_{x}} +e^{2i\theta \hat{S}_{x}}\! \right\rangle \approx N^2\theta^2-2,
\end{equation}
when $\vartheta=\pi/2+2k\pi$ and $\varphi=2k\pi$.
Through the analogous method, one can gain the maximum
\begin{equation}
-\left\langle e^{-i\theta \hat{S}_{x}}+e^{i\theta \hat{S}_{x}}\right\rangle ^{2}  \approx  N^{2}\theta^{2}-4,
\end{equation}
when $N$ is large enough, $\nu=\pi+2k\pi$, $\vartheta=\pi/2+2k\pi$ and $\varphi=2k\pi$.

Therefore, we gain the maximal QFI 
\begin{eqnarray}
\mathcal{F}_{max}  &=& 4+4\alpha^{2}\left[\frac{2\left(1-e^{-2\alpha^{2}}\cos\phi\right)}{1+e^{-2\alpha^{2}}\cos\phi}+N^{2}\theta^{2}-2\right.\nonumber\\
&& \left.+\left(N^{2}\theta^{2}-4\right)\left(\frac{e^{-2\alpha^{2}}\sin\phi}{1+e^{-2\alpha^{2}}\cos\phi}\right)^{2}\right]
\end{eqnarray}
when the relative phase factor of atomic state $\nu$, the zenith angle $\vartheta$ and the azimuthal angle $\varphi$ satisfy the optimal conditions.
With the derivative of the QFI versus the  relative phase factor $\phi$ of the optical state
\begin{eqnarray}
\partial_{\phi}\mathcal{F}_{max}\!=\!\frac{8\alpha^{2}\!\left(\!2e^{\!4\alpha^{2}\!}\!+\!e^{2\alpha^{2}}\left(\!N^{2}\theta^{2}\!-\!2\right)\!\cos\phi\!+\!N^{2}\theta^{2}\!-\!4\right)}{\left(e^{2\alpha^{2}}+\cos\phi\right)^{3}}\!\sin\phi,
\end{eqnarray}
we know the QFI $\mathcal{F}$ has extremums when $\phi=k\pi$ and the mean photon number $\alpha^2$ is large enough.
Hence, we find the maximal QFI 
\begin{equation}
\mathcal{F}_{max}\!=\!4\!+\!8\alpha^{2}\!\left[2\!+\!\left(\!N^{2}\theta^{2}\!-\!2\right)\!\coth\alpha^{2}\right]\!\approx4\!+\!8N^{2}\alpha^{2}\mu^{2}\tau^{2}
\end{equation}
and one can gain the sensitivity speeded up to attain the HL by a prefactor $N^2$.

At longer times $\tau\gtrsim1/\left(\mu\sqrt{N}\right)$, analyzing the expression of the atomic contributions of the QFI $\mathcal{F}$ (see Appendix~C for detailed derivations), we find that the odd-$N$ is superior to the even-$N$, and the maximal QFI is
\begin{equation}
\mathcal{F}_{max}\!\left(\theta\right) \! = \! \left.\mathcal{F}\right|_{\theta=\pi+2k\pi}
= \! 4\!+\!4\alpha^{2}\!\left[\!\frac{2\left(1\!-\!e^{-2\alpha^{2}}\!\cos\phi  \right)}{1\!+\!e^{-2\alpha^{2}}\cos\phi}\!+\!2\!\left(1\!+\!\cos\nu\!\cos^{N}\!\vartheta\right)\!\right]
\end{equation}
with respect to parameter $\theta$ when $\theta=\pi+2k\pi$.
When $\left.\nu+\vartheta=2k\pi\right.$ is satisfied, we have 
\begin{equation}
\mathcal{F}_{max}\left(\theta\right) 
 =  4+8\alpha^{2}\left[\frac{1-e^{-2\alpha^{2}}\cos\phi}{1+e^{-2\alpha^{2}}\cos\phi}+2\right].
\end{equation}
With $\partial\mathcal{F}_{max}=16e^{2\alpha^{2}}\alpha^{2}\sin\phi/\left(e^{2\alpha^{2}}+\cos\phi\right)^{2}$,
we know the QFI $\mathcal{F}$ is monotonically increasing when $\phi\in\left[0,\pi\right]$.
Therefore, we have the maximal QFI 
\begin{equation}
\mathcal{F}_{max}  =4+8\alpha^{2}\left(2+\coth\alpha^{2}\right)\approx 4+24\alpha^{2}
\end{equation}
when the spin number $N$ is odd, $\nu+\vartheta=2k\pi$, $\theta,\phi=\pi+2k\pi$ and $k\in\mathbb{Z}$.
It is clear that the maximal QFI is larger when we choose superposed spin-coherent states as the atomic state instead of  collective ground states.

\begin{figure}
	\centering
		\begin{minipage}{0.49\linewidth}
		\begin{overpic}[width=\linewidth]{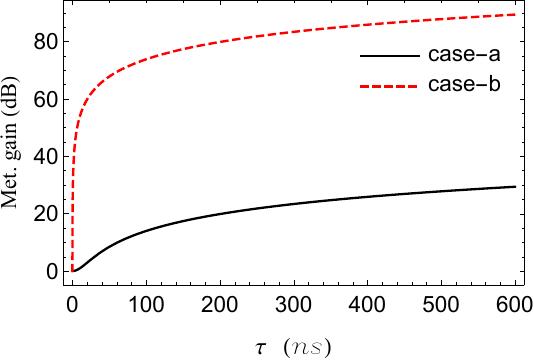}\label{fig3-a}
			\put(0,65){(a)}
		\end{overpic}
	\end{minipage}
	\begin{minipage}{0.49\linewidth}
		\begin{overpic}[width=\linewidth]{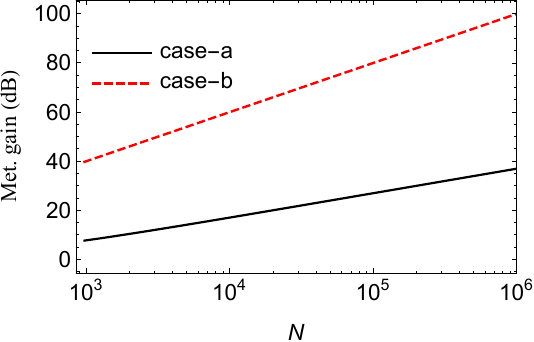}\label{fig3-b}
			\put(0,65){(b)}
		\end{overpic}
	\end{minipage}
	\protect\protect\caption{
		Metrological gains $\Delta^2 \beta_{SQL}/\left(\Delta^2 \beta\right)\equiv 1/\left(4\Delta^2 \beta\right)$ of both cases for short times $\tau\ll1/\left(\mu\sqrt{N}\right)$.
		(a) The metrological gains as functions of the evolution time $\tau$ for the case-a (black line) and the case-b (red dotted line) with the spin number fixed at $N=5\times10^5$.
		(b) The metrological gains as functions of the spin number $N$ with the evolution time fixed at $\tau=10^{-6}s$.
		The coherent amplitude $\alpha=100\sqrt{N}$, the atom-light coupling constant $\mu/\left(2\pi\alpha\right)=11\ kHz$ and the other experimental parameters could be found in~\cite{Norcia2018,Norcia2016}. 
	}
	\label{fig3}
\end{figure}

To compare the results of both cases, we calculate and show the metrological gain $\Delta^2 \beta_{SQL}/\left(\Delta^2 \beta\right)\!\equiv 1/\left(4\Delta^2 \beta\right)$ to describe the performance of our scheme for short times.
As seen in \cref{fig3}(a), the sensitivties are enhanced dramatically with the short time $\tau$ increasing. 
Moreover, the slope of the latter metrological gain is steeper than the former with respect to the spin number ($N^2$ versus $N$) [\cref{fig3}(b)].
It means that the increased spin number $N$ expedites the sensitivity to approach the HL by a prefactor $1/N^2$ in the case-b.


When it comes to decoherence due to leakage of photons $\left.\kappa\ll1/\tau\right.$, as seen in \cite{Lewis-Swan2020,barberena2020},
one can have a clear picture from the archetypal bosonic cat state
$\left|\psi\right\rangle _{cat}=\left(\left|\alpha_{1}\right\rangle +\left|\alpha_{2}\right\rangle \right)/\sqrt{2}$.
The superposition is destroyed exponentially 
\\$\exp{\left(-\kappa t\left|\alpha_{1}-\alpha_{2}\right|^{2}/2\right)}$
with the separation $\left|\alpha_{1}-\alpha_{2}\right|$ of the coherences
by photon loss.

\section{Conclusion}\label{Sec. V}
In summary, we study the time-reversal protocol to sense small displacements of the light field and corroborate the sensitivity of our scheme that can surpass the SQL and even attain the concrete HL.
Firstly, we describe the time-reversal protocol by showing its steps specifically.
Moreover, we show its unitary parametrization process and present the expression of its QFI.
Eventually, we analytically derive the QFIs for two cases with different initial states:
(a) the superposition of even and odd coherent state, and the collective ground state;
(b) the superposition of even and odd coherent state, and the superposed spin-coherent state.

We find that the QFIs of both cases approach the HL.
Furthermore, for short time, the increased spin number $N$ expedites the sensitivity to achieve the HL by a prefactor $1/N^2$ in the case-b.
However, the QFIs approach constants independent of $N$ and $t$ for both cases at longer times.
The interpretation of the result is that the atomic fluctuations will superpose the bosonic coherent state completely about a circle of radius $\left|\alpha\right|$ in phase space, and the state is equivalently sensitive to perturbations along any direction.
It is worth noting that there is no non-trivial difference between some interacting spin cases (e.g. the Ising model) and our noninteracting spin cases.
When it comes to general cases, e.g. the XY model, it is complex due to the non-commutativity.
Additionally, one may choose $\hat{M}=\hat{S}_x \cos\Phi+\hat{S}_y\sin\Phi$ ($\Phi$ is the polarized angle) as a general observble to access the maximal classical Fisher information in final measurements.
Our scheme may outline an experimentally feasible way to achieve Heisenberg-limited metrology with nonclassical states.

\section*{Acknowledgments}

This work was supported by the the National Natural Science Foundation of China (Grants No.~11935012 and No.~11875231), the National Key Research and Development Program of China (No.~2017YFA0304202 and No.~2017YFA0205700), and the Fundamental Research Funds for the Central Universities through Grant No.~2018FZA3005.

\appendix

%
%
%
%
%
%
%
%
%
%
%
%
%
\setcounter{section}{1}
\section*{Appendix A:~Derivation of \cref{F_1}}\label{Appendix_A}

Here, we derive the \cref{F_1}.
The total evolution of this system is 
\begin{eqnarray}
\hat{U}&=&e^{i\theta \hat{a}^{\dagger}\hat{a}\hat{S}_{x}}\hat{D}\left(\beta\right)e^{-i\theta \hat{a}^{\dagger}\hat{a}\hat{S}_{x}}\nonumber\\
&=& e^{i\theta \hat{a}^{\dagger}\hat{a}\hat{S}_{x}}e^{\beta \hat{a}^{\dagger}-\beta^{*}\hat{a}}e^{-i\theta \hat{a}^{\dagger}\hat{a}\hat{S}_{x}},
\end{eqnarray}
where the displacement operators satisfy
\begin{eqnarray}
\hat{D}\left(\beta\right) & =&  e^{\beta \hat{a}^{\dagger}-\beta^{*}\hat{a}},\\
\hat{D}^{\dagger}\left(\beta\right) & =&  \hat{D}\left(-\beta\right)=\hat{D}^{-1}\left(\beta\right).
\end{eqnarray}

From the Baker-Campbell-Hausdorff formula in~\cite{Baker1901}
\begin{eqnarray}
e^{\hat{X}}e^{\hat{Y}} & =& e^{\sum_{n=0}^{\infty}\frac{\left(\hat{X}^{\times}\right)^{n}}{n!}\hat{Y}}e^{\hat{X}},
\end{eqnarray}
where $\hat{X}^{\times}$ is a superoperator satisfying $\left.\hat{X}^{\times}\left(\bullet\right)=\left[\hat{X},\bullet\right]\right.$.
We take $\left.\hat{X}=i\theta \hat{a}^{\dagger} \hat{a}\hat{S}_{x}\right.$ and $\left.\hat{Y}=\beta \hat{a}^{\dagger}-\beta^{*}\hat{a}\right.$, and have
\begin{eqnarray}
\left(\hat{X}^{\times}\right)^{n}\hat{Y} & =&  \left(i\theta \hat{S}_{x}\right)^{n}\left(\beta \hat{a}^{\dagger}-\left(-1\right)^{n}\beta^{*}\hat{a}\right),\\
e^{\hat{X}}e^{\hat{Y}} & =&  e^{\sum_{n=0}^{\infty}\frac{\left(\hat{X}^{\times}\right)^{n}}{n!}\hat{Y}}e^{\hat{X}}\nonumber\\
& =&  e^{\sum_{n=0}^{\infty}\frac{\left(i\theta \hat{S}_{x}\right)^{n}\left(\beta \hat{a}^{\dagger}-\left(-1\right)^{n}\beta^{*}\hat{a}\right)}{n!}}e^{i\theta \hat{a}^{\dagger}\hat{a}\hat{S}_{x}}\nonumber\\
& = & e^{\beta e^{i\theta \hat{S}_{x}}\hat{a}^{\dagger}-\beta^{*}e^{-i\theta \hat{S}_{x}}\hat{a}}e^{i\theta \hat{a}^{\dagger}\hat{a}\hat{S}_{x}}.
\end{eqnarray}
Hence, we get the evolution operator
\begin{equation}
\hat{U}  =  e^{i\theta \hat{a}^{\dagger}\hat{a}\hat{S}_{x}}e^{\beta \hat{a}^{\dagger}-\beta^{*}\hat{a}}e^{-i\theta \hat{a}^{\dagger}\hat{a}\hat{S}_{x}}
=  e^{\beta e^{i\theta \hat{S}_{x}}\hat{a}^{\dagger}-\beta^{*}e^{-i\theta \hat{S}_{x}}\hat{a}},
\end{equation}
and the Hermitian generator with respect to the real $\beta$
\begin{eqnarray}
\hat{\mathcal{H}} & =&\! i\!\left[\partial_{\beta}e^{\!-\!\beta\!\left(\!e^{\!i\theta\! \hat{S}_{x}}\!\hat{a}^{\dagger}\!-e^{\!-\!i\theta\! \hat{S}_{x}}\!\hat{a}\!\right)\!}\!\right]\!e^{\!\beta\!\left(\!e^{i\theta\! \hat{S}_{x}}\!\hat{a}^{\dagger}\!-e^{\!-\!i\theta \hat{S}_{x}}\!\hat{a}\!\right)\!}
=\!-i\!\left(\!e^{\!i\theta \hat{S}_{x}}\!\hat{a}^{\dagger}\!-\!e^{\!-\!i\theta \hat{S}_{x}}\!\hat{a}\!\right)\!,\\
\hat{\mathcal{H}}^{2} & = & \!\left[\!-\!i\left(e^{i\theta\! \hat{S}_{x}}\!\hat{a}^{\dagger}\!-e^{\!-\!i\theta \hat{S}_{x}}\!\hat{a}\!\right)\!\right]^2
= \!2\hat{a}^{\dagger}\hat{a}\!+\!1\!-\!\left(e^{2i\theta \hat{S}_{x}}\hat{a}^{\dagger2}+e^{-2i \theta \hat{S}_{x}}\hat{a}^{2}\right).
\end{eqnarray}
Eventually, we gain the QFI of the time-reversal protocol
\begin{eqnarray}
\mathcal{F}  &=&  4\left(\left\langle \hat{\mathcal{H}}^{2}\right\rangle -\left\langle \hat{\mathcal{H}}\right\rangle ^{2}\right)\nonumber\\
& =&  4\left[2\left\langle \hat{a}^{\dagger}\hat{a}\right\rangle +1-\left(\left\langle \hat{a}^{\dagger2}\right\rangle \left\langle e^{2i\theta \hat{S}_{x}}\right\rangle +\left\langle \hat{a}^{2}\right\rangle \left\langle e^{-2i \theta \hat{S}_{x}}\right\rangle \right)\right.\nonumber\\
&&\left.+\left(\left\langle \hat{a}^{\dagger}\right\rangle \left\langle e^{i\theta \hat{S}_{x}}\right\rangle -\left\langle \hat{a}\right\rangle \left\langle e^{-i\theta \hat{S}_{x}}\right\rangle \right)^{2}\right].
\end{eqnarray}

\setcounter{section}{2}
\setcounter{equation}{0}
\section*{Appendix B:~Monotonicity of \cref{F_2}}\label{Appendix_B}

In order to analyze the monotonicity of \cref{F_2}, we first gain the derivative of the QFI versus the relative phase factor $\phi$ 
\begin{eqnarray}
\partial_{\phi}\mathcal{F} \!  &= & \!\frac{16\alpha^{2}\!\sin\phi\!\left[ e^{4\alpha^{2}}\!-\!e^{2\alpha^{2}}\left(2\!\cos^{2N}\!\frac{\theta}{2}\!-\!1\right)\!\cos\phi\!-\!2\!\cos^{2N}\!\frac{\theta}{2}\right] }{\left(e^{2\alpha^{2}}\!+\!\cos\phi\right)^{3}}\nonumber\\
&= & \frac{16\alpha^{2}\sin\phi}{\left(e^{2\alpha^{2}}+\cos\phi\right)^{3}}f,
\end{eqnarray}
where we set $\left. f= e^{4\alpha^{2}}+\left(1+e^{2\alpha^{2}} \cos\phi \right)y-1\right.$  and $y=1-2\cos^{2N}\left(\theta /2\right)$ to simplify the calculation.
Computing the derivative of $f$ with respect to the amplitude $\alpha$,
\begin{eqnarray}
\partial_{\alpha}f & = 4\alpha e^{2\alpha^{2}}\left( 2e^{2\alpha^{2}}+y \cos\phi\right), 
\end{eqnarray}
we find that the derivative $\left.\partial_{\alpha}f\geq0\right.$ and $\left.f\geq \left.f\right|_{\alpha =0}=\!\left(1\!+\!\cos\phi\right) y \geq0\right.$ for $\left.\alpha\in\left[0,\infty\right)\right.$ if $y\geq0$.

If $y\geq0$, we know that $f\geq \left.f\right|_{\alpha=0}\geq0$ and 
\begin{equation}
\partial_{\phi} \mathcal{F}  \geq\lim_{\alpha\rightarrow0}\partial_{\phi}\mathcal{F}
= \! \frac{16\alpha^{2}\sin\phi }{\left(1+\cos\phi\right)^{3}} \left.f\right|_{\alpha=0}
= \! \frac{16\alpha^{2} y \sin\phi}{\left(1+\cos\phi\right)^{2}}.
\end{equation}
Thereby, for $y\geq0$ , the QFI $\mathcal{F}$ is monotonically increasing when $\phi\in\left[2k\pi,\pi+2k\pi\right)$ and is monotonically decreasing when $\phi\in\left(\pi+2k\pi,2\pi+2k\pi\right)$, where $k\in\mathbb{Z}$.

If $y<0$, then we know the roots of the function $f$ are
\begin{eqnarray}
\!\alpha_{1}\! & =& \! \sqrt{\frac{1}{2}\!\ln\!\left\{\! \frac{1}{2}\left[\!-y\cos\!\phi\!+\!\sqrt{4\left(1-y\right)+y^2 \cos^{2}\!\phi}\right]\right\} },\\
\!\alpha_{2}\! & =& \! \sqrt{\frac{1}{2}\!\ln\!\left\{\! \frac{1}{2}\left[\!-y\cos\!\phi\!-\!\sqrt{4\left(1-y\right)+y^2 \cos^{2}\!\phi}\right]\right\} },\\
\!\alpha_{3}\! & =& \! -\!\sqrt{\frac{1}{2}\!\ln\!\left\{\! \frac{1}{2}\left[\!-y\cos\!\phi\!-\!\sqrt{\!4\!\left(1-y\right)\!+\!y^2 \cos^{2}\!\phi}\right]\right\} },\\
\!\alpha_{4}\! & =& \! -\!\sqrt{\frac{1}{2}\!\ln\!\left\{\! \frac{1}{2}\left[\!-y\cos\!\phi\!+\!\sqrt{\!4\!\left(1-y\right)\!+\!y^2 \cos^{2}\!\phi}\right]\right\} },
\end{eqnarray}
and the maximal root $\alpha_{1}\geq0$ is valid only when $y \leq 0$ is satisfied.
Therefore, when $y<0$, we find the fact as 
\begin{eqnarray}
\begin{cases}
	\partial_{\phi} \mathcal{F}\geq 0, & if\,\alpha\in\left[\alpha_{1},\infty\right)$ and \,$\phi\in\left[2k\pi,\pi+2k\pi\right),\\
	\partial_{\phi} \mathcal{F}\leq 0, & if\,\alpha\in\left[\alpha_{1},\infty\right)$ and \,$\phi\in\left[\pi+2k\pi,2\pi+2k\pi\right),\\
	\partial_{\phi} \mathcal{F}\geq 0, & if\,\alpha\in\left[0,\alpha_{1}\right)$ and \,$\phi\in\left[\pi+2k\pi,2\pi+2k\pi\right),\\
	\partial_{\phi} \mathcal{F}\leq 0, & if\,\alpha\in\left[0,\alpha_{1}\right)$ and \,$\phi\in\left[2k\pi,\pi+2k\pi\right),
\end{cases}
\end{eqnarray}
where $k\in\mathbb{Z}$.


Eventually, one can find the complete monotonicity of
the QFI, which is independent of the time-dependent phase factor $\theta$: if $y\geq0$ or $\alpha\in\left[\alpha_{1},\infty\right)$, the QFI $\mathcal{F}$ is monotonically
increasing (decreasing) when $\phi\in\left[2k\pi,\pi+2k\pi\right)$ ($\phi\in\left[\pi+2k\pi,2\pi+2k\pi\right)$); if $\left.y<0\right.$ and $\alpha\in\left[0,\alpha_{1}\right)$, the QFI $\mathcal{F}$ is monotonically increasing (decreasing) when $\phi\in\left[\pi+2k\pi,2\pi+2k\pi\right)$ ($\phi\in\left[2k\pi,\pi+2k\pi\right)$).

\setcounter{section}{3}
\setcounter{equation}{0}
\section*{Appendix C:~Derivation of atomic contributions for the superposed spin-coherent state}\label{Appendix_C}
From Eqs.~\eqref{R1} and~\eqref{SCS}, one can gain

\begin{eqnarray}
e^{-i\theta \hat{S}_{x}}\!\left|\eta\right\rangle\!  & =&  e^{-i\theta \hat{S}_{x}}\hat{R}\left(\vartheta,\varphi\right)\left|0\right\rangle \nonumber\\
& = & \hat{R}\left(-\theta,\frac{\pi}{2}\right)\hat{R}\left(\vartheta,\varphi\right)\left|0\right\rangle\nonumber\\
& = & \left(\frac{\!1\!}{1+\left|\eta\right|^{2}}\right)^{\frac{N}{2}}e^{-\xi^{*}\hat{S}\!_{-}}\!e^{\!-\!\ln\left(\!1\!+\!\left|\xi\right|^{2}\!\right)\hat{S}\!_{z}}e^{\left(\xi+\eta\right)\hat{S}\!_{+}}\!\left|0\right\rangle \nonumber\\
& = & \left(\frac{1}{1+\left|\eta\right|^{2}}\right)^{\frac{N}{2}}e^{-\xi^{*}\hat{S}_{-}}e^{-\ln\left(1+\left|\xi\right|^{2}\right)\hat{S}_{z}} \sum_{k=0}^{N}\frac{\left(\xi+\eta\right)^{k}}{k!}\sqrt{\frac{k!N!}{\left(N-k\right)!}}\left|k\right\rangle \nonumber\\
& = & \left(\frac{1}{1+\left|\eta\right|^{2}}\right)^{\frac{N}{2}}e^{-\xi^{*}\hat{S}_{-}} \sum_{k=0}^{N}\left(\xi+\eta\right)^{k}\left(\begin{array}{c}
N\\
k
\end{array}\right)^{\frac{1}{2}}e^{-\ln\left(1+\left|\xi\right|^{2}\right)\hat{S}_{z}}\left|k\right\rangle \nonumber\\
& = & \left(\frac{1}{1+\left|\eta\right|^{2}}\right)^{\frac{N}{2}}e^{-\xi^{*}\hat{S}_{-}}\sum_{k=0}^{N}\left\{\left(\xi+\eta\right)^{k}\left(\!\begin{array}{c}
\!N\!\\
\!k\!
\end{array}\!\right)^{\frac{1}{2}}\!\sum_{l=0}^{\infty}\frac{\left[-\ln\left(1+\left|\xi\right|^{2}\right)\right]^{l}}{l!}\left(\!k\!-\!\frac{N}{2}\!\right)^{l}\!\left|k\right\rangle\right\}\nonumber\\
& = &\!  \left(\frac{\!1\!+\!\left|\xi\right|^{2}\!}{\!1\!+\!\left|\eta\right|^{2}}\right)^{\!\frac{N}{2}\!}\!\sum_{k=0}^{N}{\!\left(\!\frac{\!\xi\!+\!\eta\!}{\!1\!+\!\left|\xi\right|^{2}\!}\!\right)^{k\!}\left(\begin{array}{c}
	N\\
	k
	\end{array}\right)^{\frac{1}{2}}}e^{\!-\xi^{*}\hat{S}_{-}}\!\left|k\right\rangle,
\end{eqnarray}

where the ladder operators $\hat{S}_{\pm}$ satisfy that
\begin{eqnarray}
\hat{S}^{k}_{+}\left|n\right\rangle_{\frac{N}{2}} & =& \sqrt{\frac{\left(n+k\right)!\left(N-n\right)}{n!\left(N-n-k\right)}}\left|n+k\right\rangle_{\frac{N}{2}},\\
\hat{S}^{k}_{-}\left|n\right\rangle_{\frac{N}{2}} & =& \sqrt{\frac{n!\left(N-n+k\right)}{\left(n-k\right)!\left(N-n\right)}}\left|n-k\right\rangle_{\frac{N}{2}},
\end{eqnarray}
$\eta=\frac{\varsigma}{\left|\varsigma\right|}\tan\left|\varsigma\right|=-e^{-i\varphi}\tan\frac{\vartheta}{2}$, $\varsigma=-\frac{\vartheta}{2}e^{-i\varphi}$ and $\left.\xi=-i\tan\frac{\theta}{2}\right.$.
Hence, we get the average value

\begin{eqnarray}
\left\langle \eta\right|e^{-i \theta \hat{S}_{x}}\left|\eta\right\rangle  
& = & \left(\frac{1}{1+\left|\eta\right|^{2}}\right)^{\frac{N}{2}}\left\langle 0\right|e^{\eta^{*}\hat{S}_{-}}\left(\frac{1+\left|\xi\right|^{2}}{1+\left|\eta\right|^{2}}\right)^{\frac{N}{2}}\sum_{k=0}^{N}\left(\frac{\xi+\eta}{1+\left|\xi\right|^{2}}\right)^{k}\left(\begin{array}{c}
N\\
k
\end{array}\right)^{\frac{1}{2}}e^{-\xi^{*}\hat{S}_{-}}\left|k\right\rangle \nonumber\\
& =& \left(\!1\!+\!\left|\eta\right|^{2}\!\right)^{-N}\!\left(\!1\!+\!\left|\xi\right|^{2}\right)^{\frac{N}{2}}\sum_{k=0}^{N}\left(\frac{\xi+\eta}{1+\left|\xi\right|^{2}}\right)^{k}\left(\begin{array}{c}
N\\
k
\end{array}\right)^{\frac{1}{2}}\left\langle 0 \right|e^{\left(\eta-\xi\right)^{*}\hat{S}_{-}}\left|k\right\rangle \nonumber\\
& =& \left(\!1\!+\!\left|\eta\right|^{2}\!\right)^{-N}\!\left(\!1\!+\!\left|\xi\right|^{2}\right)^{\frac{N}{2}}\nonumber\\
&& \times \sum_{k=0}^{N}\sum_{l=0}^{k}\left(\frac{\xi+\eta}{1+\left|\xi\right|^{2}}\right)^{k}\left(\begin{array}{c}
N\\
k
\end{array}\right)^{\frac{1}{2}}\left\langle 0\right|\frac{\left[\left(\eta-\xi\right)^{*}\right]^{l}}{l!}\sqrt{\frac{\left(N-k+l\right)!k!}{\left(k-l\right)!\left(N-k\right)!}}\left|k-l\right\rangle \nonumber\\
& = & \left(\!1\!+\!\left|\eta\right|^{2}\!\right)^{-N}\!\left(\!1\!+\!\left|\xi\right|^{2}\right)^{\frac{N}{2}}\nonumber\\
&& \times \sum_{k=0}^{N}\sum_{l=0}^{k}\left(\frac{\xi+\eta}{1+\left|\xi\right|^{2}}\right)^{k}\left(\begin{array}{c}
N\\
k
\end{array}\right)^{\frac{1}{2}}\left(\begin{array}{c}
N-k+l\\
l
\end{array}\right)^{\frac{1}{2}}\left(\begin{array}{c}
k\\
l
\end{array}\right)^{\frac{1}{2}}\left[\left(\eta-\xi\right)^{*}\right]^{l}\delta_{k,l}\nonumber\\
& = & \left(\!1\!+\!\left|\eta\right|^{2}\!\right)^{-N}\!\left(\!1\!+\!\left|\xi\right|^{2}\right)^{\frac{N}{2}}\sum_{k=0}^{N}\left(\frac{\left(\eta-\xi\right)^{*}\left(\eta+\xi\right)}{1+\left|\xi\right|^{2}}\right)^{k}\left(\begin{array}{c}
N\\
k
\end{array}\right)\nonumber\\
& = & \cos^{2N}\left(\frac{\vartheta}{2}\right)\cos^{-N}\left(\frac{\theta}{2}\right)\left\{ \cos^{2}\left(\frac{\theta}{2}\right)\left[\sec^{2}\left(\frac{\vartheta}{2}\right)+2i\tan\left(\frac{\theta}{2}\right)\tan\left(\frac{\vartheta}{2}\right)\cos\varphi\right]\right\} ^{N}\nonumber\\
& =&  \left[\cos\left(\frac{\theta}{2}\right)+i\sin\left(\frac{\theta}{2}\right)\sin\vartheta \cos\varphi\right]^{N}.
\end{eqnarray}

Through the analogous calculations, one can obtian that
\begin{eqnarray}
\left\langle \eta\right|\!e^{\pm 2i\lambda\theta \hat{S}_{\!x\!}}\!\left|\eta\right\rangle \!  & = &\!  \left[\cos\!\left(\lambda\theta \right)\!\mp\!i\sin\!\left(\lambda\theta \right)\!\sin \! \vartheta \! \cos \! \varphi \right]^{N\!}\!,\\
\left\langle \!-\eta\right|\!e^{\pm 2i\lambda\theta \hat{S}_{\!x\!}}\!\left| -\eta\right\rangle \!  & =& \!  \left[\cos\!\left(\lambda\theta \right)\!\pm\!i\sin\!\left(\lambda\theta \right)\!\sin \! \vartheta \! \cos \! \varphi \right]^{N\!}\!,\\
\left\langle \eta\right|\!e^{\pm 2i\lambda\theta \hat{S}_{\!x\!}}\!\left| -\eta\right\rangle \!  & = &\!  \left[\cos\!\left(\lambda\theta \right)\!\cos\!\vartheta\!\pm\!\sin\!\left(\lambda\theta \right)\!\sin \! \vartheta \! \sin \! \varphi \right]^{N\!}\!,
\end{eqnarray}
where $\lambda=1/2,\,1$ is a constant.

To compute the atomic contributions, we set 
\begin{equation}
\left\langle \hat{O}\right\rangle  =  \frac{1}{2}\left(\left\langle \eta\right|\hat{O}\left|\eta\right\rangle +\left\langle -\eta\right|\hat{O}\left|-\eta\right\rangle +e^{i\nu}\left\langle \eta\right|\hat{O}\left|-\eta\right\rangle +e^{-i\nu}\left\langle -\eta\right|\hat{O}\left|\eta\right\rangle \right)
\end{equation}
and eventually gain 
\begin{equation}
\left\langle \! e^{\!-i\zeta\theta \hat{S}_{x}}\! +\!e^{i\zeta\theta \hat{S}_{x}}\! \right\rangle \!  \! = \! \left\langle \eta\right|\!e^{-i\zeta\theta \hat{S}_{\!x\!}}\!+\!e^{i\zeta\theta \hat{S}_{\!x\!}}\!\left|\eta\right\rangle
\!\quad \!+\!\cos \nu \! \left(\!\left\langle \eta\right|\!e^{-i\zeta \theta \hat{S}_{\!x\!}}\!+\!e^{i\zeta \theta \hat{S}_{\!x\!}}\!\left| -\eta\right\rangle\!\right),
\end{equation}
where $\zeta=1,\,2$ is a constant.

\section*{References}
\bibliographystyle{plain}
\bibliography{TRP_citebook}

\end{document}